\documentstyle[aps,preprint,epsf,axodraw]{revtex}      
\newcommand{\postscript}[2] {\setlength{\epsfxsize}{#2\hsize}
\centerline{\epsfbox{#1}}}

\tightenlines
\begin{document}
\title{Relativistic Quark Spin Coupling Effects in the Nucleon
Electromagnetic Form Factors}
\author{ W.R.B. de Ara\'ujo$^a$, E.F. Suisso$^b$, 
T. Frederico$^b$, M. Beyer$^c$, and
  H.J. Weber$^c$\footnote{permanent address: Dept. of Physics,
University of Virginia, Charlottesville, U.S.A.}}
\address{$^a$Laborat\'orio do Acelerador Linear, Instituto de F\'\i sica
da USP\\ C.P. 663118, CEP 05315-970, S\~ao Paulo, Brazil}
\address{$^b$ Dep. de F\'\i sica, Instituto Tecnol\'ogico de Aeron\'autica,
Centro T\'ecnico Aeroespacial, \\
12.228-900 S\~ao Jos\'e dos Campos, S\~ao Paulo, Brazil.}
\address{$^c$ FB Physik, Universit\"at Rostock, 18051 Rostock, Germany}
\maketitle
\begin{abstract}
  We investigate the effect of different forms of relativistic spin
  coupling of constituent quarks in the nucleon electromagnetic form
  factors. The four-dimensional integrations in the two-loop Feynman
  diagram are reduced to the null-plane, such that the light-front
  wave function is introduced in the computation of the form factors.
  The neutron charge form factor is very sensitive to different
  choices of spin coupling schemes, once its magnetic moment is fitted
  to the experimental value. The scalar coupling between two quarks is
  preferred by the neutron data, when a reasonable fit of the proton
  magnetic momentum is found.

\end{abstract}

\vspace{3ex}

The purpose of this work is to study the nucleon
electromagnetic form factors using  different forms of
relativistic spin coupling between the constituent
quarks forming the nucleon. We use an effective Lagrangian to describe
the quark spin coupling to the nucleon.
We keep close contact with covariant field theory and perform
a three-dimensional reduction of the amplitude
for the photon absorption process by the nucleon
to the null-plane, $x^+=x^0+x^3=0$,
 (see, e.g., Ref.~\cite{karmanov}). After the
three-dimensional reduction, one can introduce
the nucleon light-front wave function in the two-loop
momentum integrations which define the electromagnetic current.
We consider the triangle diagram that is a major ingredient of recent
electromagnetic and weak baryon form factor evaluations in light front
dynamics.

We start with the effective Lagrangian for the N-q coupling
\begin{eqnarray}
{\cal{L}}_{N-3q}=\alpha m_N
\epsilon^{ijk}
\overline{\Psi}_{(i)}
i\tau _2\gamma _5\Psi_{(j)}^C\overline{\Psi} _{(k)}\Psi _N
+(1-\alpha) 
\epsilon^{ijk}
\overline{\Psi}_{(i)}
i\tau _2\gamma_\mu \gamma _5\Psi_{(j)}^C
\overline{\Psi} _{(k)}i\partial^\mu\Psi _N
+ H.C.
\label{lag}
\end{eqnarray}
where $\tau _2$ is the isospin matrix, 
the color indices are $\{i,j,k\}$ and  $\epsilon^{ijk}$ is the
totally antisymmetric symbol.
The conjugate quark field is $\Psi^C=C
\overline{\Psi}^\top $, where $C=i\gamma^2\gamma^0$ is the charge
conjugation matrix; $\alpha$ is a parameter to dial the spin
coupling parameterization, and $m_N$ is the nucleon mass.

The macroscopic matrix element of the nucleon electromagnetic
current $j^+_N(Q^2)$ in the Breit-frame and in the light-front
spinor basis is given by:
\begin{eqnarray}
\langle s'|j^+_N(Q^2)|s\rangle &=&\bar{u}(p',s')
\left( F_{1N}(Q^2)\gamma^++ i\frac{\sigma^{+\mu}Q_\mu}{2 m_N}F_{2N}(Q^2)
\right) {u}(p,s)
\nonumber \\
&=& \frac{p^+}{m_N}
\langle s'| F_{1N}(Q^2)+ i\frac{F_{2N}(Q^2)}{2 m_N}
\vec n\cdot (\vec q_\perp \times \vec \sigma )| s \rangle \ ,
\label{jp}
\end{eqnarray}
where $F_{1N}$ and $F_{2N}$ are the Dirac and Pauli
form factors, respectively. $\vec n$ is the unit vector along the
z-direction. The Breit-frame momenta are $Q^\mu=(0,\vec q_\perp)$,
such that $(Q^+=Q^0+Q^3=0)$ and $\vec q_\perp=(q^1,q^2)$; 
$p=(\sqrt{\frac{q^2_\perp}{4}+m^2_N},-\frac{\vec q_\perp}{2})$ and 
$p'=(\sqrt{\frac{q^2_\perp}{4}+m^2_N},\frac{\vec q_\perp}{2})$. 

The light-front spinors are:
\begin{eqnarray}
u(p,s) =\frac{\rlap\slash p+ m_N}{2\sqrt{p^+m_N}}\gamma^+
\gamma^0 \left(\begin{array}{c}
\chi^{\rm Pauli}_{s} \\0\end{array}
\right)
\ .
\label{lf}
\end{eqnarray}
The Dirac spinor of the instant form
\begin{eqnarray}
u_{D}(p,s)=\frac{\rlap\slash p
+m_N}{\sqrt{2 m(p^0+m) }}
\left(\begin{array}{c}
\chi^{\rm Pauli}_{s} \\
0
\end{array}
\right)  \
\label{dirac}
\end{eqnarray}
carries the subscript $D$.

The Sachs form factors are defined by:
\begin{eqnarray}
G_{EN}(Q^2)&=& F_{1N}(Q^2)+\frac{Q^2}{4m_N^2}F_{2N}(Q^2) \ ,
\nonumber \\
G_{MN}(Q^2)&=& F_{1N}(Q^2)+F_{2N}(Q^2) \ .
\end{eqnarray}
The magnetic moment is $\mu_N=G_{MN}(0)$ and the mean square
radius is $r^2_N=6\frac{dG_{EN}(Q^2)}{dQ^2}|_{Q^2=0}$.

The nucleon electromagnetic current ($j_N^{+}(Q^2)$), obtained
from the effective Lagrangian by considering the complete
antisymmetrization of the matrix element of the current, has five
topologically distinct diagrams. We can calculate the photon
absorption amplitude considering only the process on quark 3,
due to the symmetrization of the
microscopic matrix element after the factorization of the
color degree of freedom. Figure (1a) defines the
nucleon spin-space operator $j^+_{aN}$ and
represents the case where quark 3 absorbs
the photon while 1 and 2 compose the spectator-coupled quark pair in
Eq.(\ref{lag}) for the initial and final nucleons. In Figure (1b), the
coupled quark pair in the initial nucleon is (13) and in the
final nucleon the coupled  pair is (12). The operator $j^+_{bN}$,
represented by the diagram (1b), should be multiplied by a factor of 4.
A factor 2 comes from the exchange of
quarks 1 and 2, and  another factor  2 comes from the invariance of
this term under the exchange of the pairs in
the initial and final  nucleons, which is a consequence of time reversal
invariance and parity transformation property. The operator
$j^+_{cN}$ is represented by  figure (1c), where the initial coupled pair
quark is (13) and the final coupled pair is (23). This operator is multiplied
by a factor of 2 because the quarks 1 and 2 can be exchanged. The operator
represented by diagram (1d), $j^+_{dN}$, corresponds to the process in which
the photon is absorbed by the coupled quark pair (23) while 1 is spectator.
In this case, two diagrams are possible by
the exchange of quarks 1 and 2. Thus, the microscopic
operator of the nucleon current is given by the sum of four terms:
\begin{eqnarray}
j^+_N(Q^2)=j^+_{aN}(Q^2)+4j^+_{bN}(Q^2)+2j^+_{cN}(Q^2)+2j^+_{dN}(Q^2)
\ .
\label{mjp}
\end{eqnarray}

The nucleon current operators $j^+_{\beta N}$, $\beta=a,b,c,d$, are written
directly from the Feynman diagrams of fig.1.
The electromagnetic field is coupled in the usual minimal way 
ensuring gauge invariance.
The electromagnetic current, $j^+_{N}$, is constructed from
the Feynman triangle two-loop diagrams of figures (1a) to (1d):
\begin{eqnarray}
\langle s'|j^+_{a N}(Q^2)|s\rangle &=& -
\langle N|\hat Q_q| N\rangle Tr[i\tau_2(-i)\tau_2]
\int \frac{d^4k_1d^4k_2}{(2\pi)^8}\Lambda(k_i,p^{\prime})\Lambda(k_i,p)
\bar u(p',s')S(k'_3)\gamma^+
\nonumber \\
&\times& S(k_3)u(p,s)
Tr\left[S(k_2)\left(\alpha m_N+(1-\alpha)\rlap\slash p\right)
\gamma^5 S_c(k_1)\gamma^5\left(\alpha m_N+(1-\alpha)\rlap\slash p'\right)
\right]
\ ,  \label{j+a}
\end{eqnarray}
with $\displaystyle S(p)=\frac{1}{\rlap\slash p-m+\imath \epsilon} \, ,$
and $\displaystyle S_c(p)=\left[\gamma^0\gamma^2
\frac{1}{\rlap\slash p-m+\imath \epsilon}\gamma^0\gamma^2\right]^\top \, $.
$m$ is the constituent quark mass and $k'_3=k_3+Q \, $.
The quark charge operator $\hat Q_q$ is diagonal and its matrix elements
are 2/3 for the up quark  and -1/3 for the down quark.
The choice of the function $\displaystyle \Lambda(k_i,p)$ will be discussed
later.

The diagram of figure (1b) is given by:
\begin{eqnarray}
\langle s'|j^+_{b N}(Q^2)|s\rangle &=& - \langle N|\hat Q_q| N\rangle
\int \frac{d^4k_1d^4k_2}{(2\pi)^8}\Lambda(k_i,p^{\prime})
\Lambda(k_i,p)
\bar u(p',s')S(k'_3)\gamma^+S(k_3)
\nonumber \\
&\times& \left(\alpha m_N+(1-\alpha)\rlap\slash p\right)\gamma^5
S_c(k_1)\gamma^5\left(\alpha m_N+(1-\alpha)\rlap\slash p'\right)
 S(k_2)u(p,s)
\ .  \label{j+b}
\end{eqnarray}

The diagram of figure (1c) is given by:
\begin{eqnarray}
\langle s'|j^+_{c N}(Q^2)|s\rangle &=& \langle N|\hat Q_q| N\rangle
\int \frac{d^4k_1d^4k_2}{(2\pi)^8}\Lambda(k_i,p^{\prime})
\Lambda(k_i,p) \bar u(p',s')S(k_1)
\left(\alpha m_N+(1-\alpha)\rlap\slash p\right)
\nonumber \\
&\times&
\gamma^5 S_c(k_3)\gamma^+
S_c(k'_3)\gamma^5\left(\alpha m_N+(1-\alpha)\rlap\slash p'\right)
S(k_2)u(p,s)
\ .  \label{j+c}
\end{eqnarray}

The diagram of figure (1d) is given by:
\begin{eqnarray}
\langle s'|j^+_{d N}(Q^2)|s\rangle &=&-
Tr[\hat Q_q]\int \frac{d^4k_1d^4k_2}{(2\pi)^8}\Lambda(k_i,p^{\prime})
\Lambda(k_i,p)\bar u(p',s')S(k_2)u(p,s)
\nonumber \\
&&Tr\left[
\gamma^5\left(\alpha m_N+(1-\alpha)\rlap\slash p'\right)S(k'_3)\gamma^+
S(k_3)\left(\alpha m_N+(1-\alpha)\rlap\slash p\right)
\gamma^5 S_c(k_1)\right]
\ .  \label{j+d}
\end{eqnarray}

The light-front coordinates are defined as 
$k^+=k^0+k^3\ , k^-=k^0-k^3 \ , k_\perp=(k^1,k^2).$
In each term of the nucleon current, from $j^+_{aN}$ to $j^+_{dN}$,
the Cauchy integrations over  $k^-_1$ and $k^-_2$ are performed.
That means the
on-mass-shell pole of the Feynman propagators for the spectator particles
1 and 2, in the photon absorption process, are taken into account.
In the Breit-frame, with $Q^+=0$, there is a maximal suppression of
light-front Z-diagrams in $j^+$ \cite{tob92,pach99}.
Thus, the components of the momentum
$k^+_1$ and $k^+_2$ are bounded, such that $ 0< k^+_1 < p^+$ and
$0<k^+_2 <p^+-k^+_1$\cite{pach97}. The four-dimensional integrations
of Eqs.(\ref{j+a}) to (\ref{j+d}) are reduced to the three-dimensional
volume of the null-plane.

The analytical integration of Eq.(\ref{j+a}) of the `-' components of
momentum yields:
\begin{eqnarray}
\langle s'|j^+_{a N}(Q^2)|s\rangle  &=& 2p^{+2}
\langle N|\hat Q_q|N\rangle
\int \frac{d^{2} k_{1\perp} dk^{+}_1d^{2} k_{2\perp} d
k^{+}_2 }{4(2\pi)^6k^+_1k^+_2k^{+\ 2}_3} \theta(p^+-k^+_1)
\theta(p^+-k^+_1-k^+_2) \nonumber \\
&&Tr\left[ (\rlap\slash k_2+m)\left(\alpha m_N+(1-\alpha)\rlap\slash p\right)
(\rlap\slash k_1+m)\left(\alpha m_N+(1-\alpha)\rlap\slash p'\right)\right]
\nonumber \\
&&\bar u(p',s')(\rlap\slash k'_3+m))\gamma^+(\rlap\slash k_3+m)u(p,s)
\frac{\Lambda(k_i,p^{\prime})}{m^2_N-M^{'2}_0}
\frac{\Lambda(k_i,p)}{m^2_N-M^2_0}
 \ ,
\label{j+alf}
\end{eqnarray}
where $k^2_1=m^2$ and $k^2_2=m^2$ and the free three-quark squared mass
is defined by:
\begin{equation}
M^2_0=p^+(\frac{k_{1\perp}^{2}+m^2}{k^+_1}+\frac{k_{2\perp}^{2}+m^2}{k^+_2}
+\frac{k_{3\perp}^{2}+m^2}{k^+_3})-{p^2_\perp} \ ,
\end{equation}
and $M^{\prime 2}_0=M^2_0(k_3\rightarrow k'_3 \ , \vec p_\perp\rightarrow
\vec p^\prime_\perp)$.

The other terms of the nucleon current, as given by Eqs.
(\ref{j+b})-(\ref{j+d}) are also integrated over the $k^-$ momentum components
of  particles 1 and 2, following the same steps used to
obtain Eq.(\ref{j+alf}) from Eq.(\ref{j+a}):
\begin{eqnarray}
\langle s'|j^+_{b N}(Q^2)|s\rangle  &=& p^{+2}
\langle N|\hat Q_q| N\rangle
\int \frac{d^{2} k_{1\perp} dk^{+}_1d^{2} k_{2\perp} d
k^{+}_2 }{4(2\pi)^6k^+_1k^+_2k^{+\ 2}_3} \theta(p^+-k^+_1)
\theta(p^+-k^+_1-k^+_2) \nonumber \\
&&\bar u(p',s')(\rlap\slash k'_3+m)\gamma^+(\rlap\slash k_3+m)
\left(\alpha m_N+(1-\alpha)\rlap\slash p\right) (\rlap\slash k_1+m)
\nonumber \\
&&\times \left(\alpha m_N+(1-\alpha)\rlap\slash p'\right)
 (\rlap\slash k_2+m)u(p,s)
\frac{\Lambda(k_i,p^{\prime})}{m^2_N-M^{'2}_0}
\frac{\Lambda(k_i,p)}{m^2_N-M^2_0}
\ ,  \label{j+blf}
\end{eqnarray}
\begin{eqnarray}
\langle s'|j^+_{c N}(Q^2)|s\rangle  &=&p^{+2}\langle N|\hat Q_q| N\rangle
\int \frac{d^{2} k_{1\perp} dk^{+}_1d^{2} k_{2\perp} d
k^{+}_2 }{4(2\pi)^6k^+_1k^+_2k^{+\ 2}_3} \theta(p^+-k^+_1)
\theta(p^+-k^+_1-k^+_2) \nonumber \\
&&\bar u(p',s')(\rlap\slash k_1+m)
\left(\alpha m_N+(1-\alpha)\rlap\slash p\right)
(\rlap\slash k_3+m)\gamma^+
(\rlap\slash k'_3+m)
\nonumber \\
&&
\times \left(\alpha m_N+(1-\alpha)\rlap\slash p'\right)
 (\rlap\slash k_2+m)u(p,s)
\frac{\Lambda(k_i,p^{\prime})}{m^2_N-M^{'2}_0}
\frac{\Lambda(k_i,p)}{m^2_N-M^2_0}
\ ,  \label{j+clf}
\end{eqnarray}
\begin{eqnarray}
\langle s'|j^+_{d N}(Q^2)|s\rangle  &=& p^{+2}Tr[\hat Q_q]
\int \frac{d^{2} k_{1\perp} dk^{+}_1d^{2} k_{2\perp} d
k^{+}_2 }{4(2\pi)^6k^+_1k^+_2k^{+\ 2}_3} \theta(p^+-k^+_1)
\theta(p^+-k^+_1-k^+_2)
\nonumber \\
&&Tr\left[
\left(\alpha m_N+(1-\alpha)\rlap\slash p'\right)(\rlap\slash k'_3+m)
\gamma^+ (\rlap\slash k_3+m)\left(\alpha m_N+(1-\alpha)\rlap\slash p\right)
(\rlap\slash k_1+m)\right]
\nonumber \\
&&\bar u(p',s')(\rlap\slash k_2+m)u(p,s)
\frac{\Lambda(k_i,p^{\prime})}{m^2_N-M^{'2}_0}
\frac{\Lambda(k_i,p)}{m^2_N-M^2_0}
\ .  \label{j+dlf}
\end{eqnarray}

The Gaussian wave function is introduced in Eqs.(\ref{j+alf})-(\ref{j+dlf}) 
through the substitution \cite{tob92}:
\begin{eqnarray}
\frac{1}{2(2\pi)^3}
\frac{\Lambda(k_i,p)}{m^2_N-M^2_0}\rightarrow N\ \exp(-M^2_0/2\beta^2)
\ ,
\end{eqnarray}
where $N$ is chosen such that the proton charge is 1.

The spin-flavor invariant of the nucleon with quark pair spin zero
($\alpha =1$) is
the simplest of a basis of 8 such states given in greater detail in
Ref.~\cite{BKW98}, for example. The only nucleon spin invariant used
and tested in form factor calculations contains the additional
projector $\rlap\slash p +M_0$ onto large Dirac components, a
characteristic feature of the Bakamjian-Thomas formalism
\cite{BT53}.
The spin-flavor invariant in the effective Lagrangian, (\ref{lag}),
with $\alpha=1/2$, resembles the Bakamjian-Thomas, but is not
 equivalent to it, as it will be explained below.

As we have discussed, the residues of the triangle Feynman diagram are
evaluated at the
on-$k^-$-shell poles of the spectator particles~\cite{pach97}. The
numerator of the fermion propagator of the quark which absorbs the
photon momentum can be considered on-$k^-$-shell because
$(\gamma^+)^2=0$. Thus,
all the numerators of the fermion propagators can be substituted by
the positive energy spinor projector, written in terms of light-front
spinors. To explore the physical meaning of the effective Lagrangian,
we follow closely the work of Ref.\cite{araujo99}, where the case 
 $\alpha=1$ is discussed in detail. It has been suggested to perform
a kinematical light-front boost of the matrix
elements of the spin operators between quark states and quark-nucleon states,
related to the initial and final nucleons and their respective rest-frames.
Because the Wigner rotation is unity for such Lorentz transformations, the
matrix elements appearing in the nucleon electromagnetic current, corresponding
to the spin coupling of the quarks to the initial or final nucleon, can be
evaluated in their respective rest-frames.
A typical matrix element of the spin coupling coefficient for $\alpha=1$
 appearing in the evaluation of
$j^+$, when calculated in the nucleon rest frame,
is given by:
\begin{equation}
\chi (s_1,s_2,s_3;s_N)=\overline{u}_{1}
\gamma _5 u_{2}^C\;
\overline{u}_{3}u_{N} \ ,
\label{nuc}
\end{equation}
where  $u_i=u(k_i,s_i)$ is the light-front spinor for the
$i$-th quark.

To calculate Eq.~(\ref{nuc}), we begin by evaluating
the matrix element of the pair coupled to spin zero in
the rest frame of the pair (c.m.) which, again, is found by a
kinematical light-front boost from the nucleon rest frame. Because the
Wigner rotation is unity for such a Lorentz transformation, we can
write (viz. ${u}_{\rm c.m.}(\vec k^{{\rm c.m.}},s) = {u}(\vec
k^{{\rm c.m.}},s)$):
\begin{eqnarray}
I(s_1,s_2,0)&=&\overline{u}(\vec k_1,s_1)
\gamma _5u^C(\vec k_2,s_2)\nonumber\\
&=& \overline{u}(\vec k^{{\rm c.m.}}_1,s_1)
\gamma_5 u^C(\vec k^{{\rm c.m.}}_2,s_2)
\ , \label{eqn:i}
\end{eqnarray}
where $\vec k^{{\rm c.m.}}=(k^{+{\rm c.m.}},{\vec k}_\perp^{{\rm c.m.}})$ are
the kinematical momentum variables of each particle 1 or 2 in the rest
frame of the pair 12, $k^{{\rm (c.m.)}\mu}=(\Lambda k)^\mu$.
The particle momenta in the pair rest frame are
obtained by a kinematical light-front transformation from those in the
nucleon rest frame to the pair rest frame.
Introducing the completeness
relation for positive energy Dirac spinors in Eq.~(\ref{eqn:i}), one finds:
\begin{eqnarray}
I(s_1,s_2,0)&=&\sum_{\bar s_1 \bar s_2}
\overline{u}(\vec k^{{\rm c.m.}}_1,s_1)u_D(\vec k^{{\rm c.m.}}_1,\bar s_1)
\overline u_D(\vec k^{{\rm c.m.}}_1,\bar s_1)\nonumber\\
&&\gamma _5\, C\, \overline{u}^\top_D(\vec k^{{\rm c.m.}}_2,\bar s_2)
\left(\overline{u}(\vec k^{{\rm c.m.}}_2,s_2)
u_D(\vec k^{{\rm c.m.}}_2,\bar s_2)\right)^\top
\ . \label{eqn:i2}
\end{eqnarray}
The Clebsch-Gordan coefficients are found by using
the Dirac spinors in Eq.(\ref{eqn:i2})
\begin{eqnarray}
\overline u_D(\vec k^{{\rm c.m.}}_1,\bar s_1)\gamma _5 C
\overline{u}^\top_D(\vec k^{{\rm c.m.}}_2,\bar s_2)
\rightarrow\chi_{\bar s_1}^\dagger i\sigma_2\chi^*_{\bar s_2}
=\sqrt{2}\;\langle \frac{1}{2} \bar s_1  \frac{1}{2} \bar s_2|
00\rangle \ . \label{i3}
\end{eqnarray}

The Melosh rotation is given by:
\begin{eqnarray}
\left[R_{M}(p)\right]_{s's}=\overline{u}_D(p,s')u(p,s)
\ .
\label{mel}
\end{eqnarray}

{}From Eqs.(\ref{nuc}),  (\ref{eqn:i2}), (\ref{i3}), and (\ref{mel}),
the form of the spin coupling of the nucleon to the quarks is written
as:
\begin{equation}
\chi (s_1,s_2,s_3;s_N)= \sum_{\bar s_1 \bar s_2}
\left[ R^\dagger_M(\vec k_1^{{\rm c.m.}})\right]_{s_1\bar s_1}
\left[ R^\dagger_M(\vec k_2^{{\rm c.m.}})\right]_{s_2\bar s_2}
\left[ R^\dagger_M(\vec k_3)\right]_{s_3 s_N}
\chi_{\bar s_1}^\dagger i\sigma_2\chi^*_{\bar s_2} \;
. \label{numel}
\end{equation}
The Melosh rotations of
the spin-zero coupled pair (12) have the momentum arguments evaluated
in the rest frame of the pair in Eq.~(\ref{numel}), while in the BT
construction the arguments of the Melosh rotations are all evaluated
in the nucleon rest frame. Moreover, the light-front spinors in
Eqs.(\ref{eqn:i}),(\ref{eqn:i2}) are no longer all in the nucleon rest
frame, which will have observable consequences in the electromagnetic
form factors. Also, the various total momentum $'+'$ components, such as
$p^+_{12}$ and $p^+$ now appear in different frames, whereas in the BT
case only $M_0$ occurs for $p^+$ in the nucleon rest frame.

The same considerations will apply to the pair-spin 0 invariant with
an additional $\rlap\slash  p+m_N$ ($\alpha=1/2$)
from the projector which reduces to
$\gamma_0+1$ in the nucleon rest frame. However, the Melosh rotations
have as arguments the quark momenta in the nucleon rest-frame. This last
case still differs from the BT construction because the sum
of the $'+'$ components of the  quark momenta is the nucleon momentum, $p^+$,
and not $M_0$ as in the BT formalism.  After this formal discussion,
in  next we discuss the observable effects of different choices
of quark spin coupling schemes for the nucleon electromagnetic current.

The numerical results for the nucleon electromagnetic form-factors for
$\alpha=$ 1, 0 and 1/2 are discussed in the following. The quark mass
is 220 MeV \cite{salme}. In our calculations the size parameter
$\beta$ is adjusted such that the neutron magnetic moment becomes
-1.91 $\mu_{\rm N}$. This choice of parameterization through the
neutron magnetic moment was used in order to focus our study on the
electric neutron form factor, which turns out to be very sensitive to
different spin coupling schemes. In Table I, the results are shown for
the nucleon mean square radius and proton magnetic moment. The neutron
mean square radius depends strongly on the spin coupling parameter
$\alpha$. The best result for the neutron mean square radius is found
for $\alpha=1$, i.e., the scalar coupling of the quark pair. The
proton observables are just consistent with the data. However, a
simultaneous fit of the nucleon magnetic moments and mean square radii
does not seem possible with good precision.

In figure 2, we present the model results for the neutron electric form-factor
$G_{En}$, and the scalar coupling is shown to be consistent with data. The
calculation with $\alpha=1/2$ is below the data as anticipated from the small 
radius. The calculation with $\alpha=0$ yields a negative $G_{En}$ and it 
was not shown in the figure. 
The model results for  $G_{Mn}$
is compared to the data in figure 3. Below momentum transfers 
of 1 GeV/c, the model is consistent with the experimental data and
above  this value it deviates from the data. The difference
between the model and experimental values of $G_{Mn}$ could be minimized, 
at the expense of increasing the deviation of the 
neutron mean square radius from its experimental value. 

In figure 4, we show the ratio of the proton electric to magnetic
form factors compared with recent data.  The choice of $\alpha=1$,
among the values of 1, 1/2 and 0, shows the smallest deviation from the data.
 However, the calculation still misses the experimental values 
which could be finally fitted introducing constituent quarks
form factors \cite{salme}, but this is beyond our purpose in this work.

In conclusion, we have tested different spin coupling schemes for
the nucleon in a calculation of the nucleon electromagnetic
form factors. We find that the electric form factor of the neutron
can be used to constrain the quark spin coupling schemes.
The comparison with the neutron data below momentum transfer of 1 GeV/c
suggests that the scalar pair is preferred in the relativistic quark
spin coupling of the nucleon.

{\bf Acknowledgments:}  HJW is grateful to G. R\"opke and the many
particle research group for their warm hospitality. 
WRBA thanks CNPq for  financial support and LCCA/USP for providing 
computational facilities, EFS thanks FAPESP for
financial support and TF thanks 
CNPq, FAPESP and CAPES/DAAD/PROBRAL.

\begin{table}
\caption{ Nucleon low-energy electromagnetic observables 
for different spin coupling parameters with a
gaussian light-front wave function adjusted to
$\mu_n=-1.91 \mu_{\rm N}$.}
\vspace{1 cm}
\begin{tabular}{|c|c|c|c|c|}
$\alpha$&$\beta$(MeV)&$<r^2>_n({\rm fm}^2)$
& $<r^2>_p({\rm fm}^2)$ &  $\mu_p(\mu_{\rm N})$\\ \hline
1       & 562  &  -0.075 & 0.877  & 3.11 \\
1/2     & 664  &  -0.024 & 0.718  & 3.05  \\
0       & 661  &   0.079 & 0.686  & 2.90 \\
Exp.    & - &  -0.113 $\pm$ 0.005\cite{kop}& 0.66 $\pm$ 0.06\cite{brod} ,
0.74 $\pm$ 0.02 \cite{mur}& 2.79 \\
\end{tabular}
\end{table}
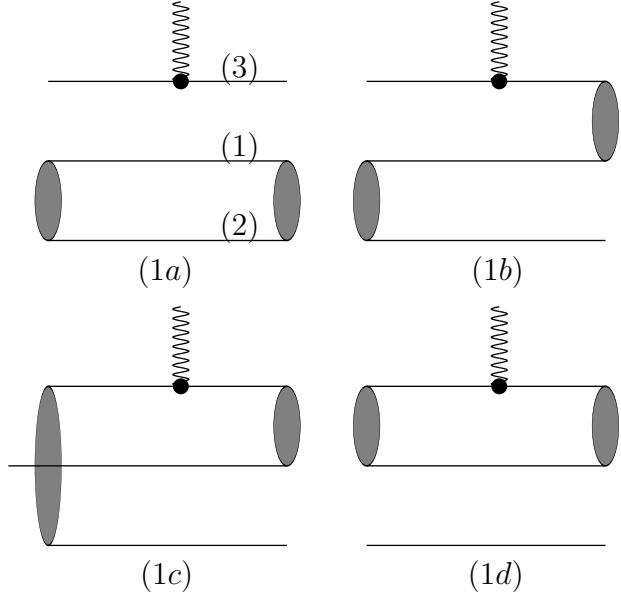
\begin{figure}[h]
\begin{center}
\vspace{5cm}
\centerline{\
\begin{picture}(330,130)(-5,-130)
\GOval(30,25)(15,5)(0){.5}
\GOval(120,25)(15,5)(0){.5}
\Photon(80,70)(80,100){3}{8.5}
\Vertex(80,70){3.0}
\put(110,12){\makebox(0,0)[br]{$(2)$}}
\put(110,42){\makebox(0,0)[br]{$(1)$}}
\put(110,72){\makebox(0,0)[br]{$(3)$}}
\Line(30,10)(120,10)
\Line(30,40)(120,40)
\Line(30,70)(120,70)
\put(85,-5){\makebox(0,0)[br]{$(1a)$}}
\GOval(150,25)(15,5)(0){.5}
\GOval(240,55)(15,5)(0){.5}
\Photon(200,70)(200,100){3}{8.5}
\Vertex(200,70){3.0}
\Line(150,10)(240,10)
\Line(150,40)(240,40)
\Line(150,70)(240,70)
\put(210,-5){\makebox(0,0)[br]{$(1b)$}}
\GOval(30,-75)(30,5)(0){.5}
\GOval(120,-60)(15,5)(0){.5}
\Photon(80,-45)(80,-15){3}{8.5}
\Vertex(80,-45){3.0}
\Line(30,-105)(120,-105)
\Line(15,-75)(120,-75)
\Line(30,-45)(120,-45)
\put(85,-120){\makebox(0,0)[br]{$(1c)$}}
\GOval(150,-60)(15,5)(0){.5}
\GOval(240,-60)(15,5)(0){.5}
\Photon(200,-45)(200,-15){3}{8.5}
\Vertex(200,-45){3.0}
\Line(150,-105)(240,-105)
\Line(150,-75)(240,-75)
\Line(150,-45)(240,-45)
\put(210,-120){\makebox(0,0)[br]{$(1d)$}}
\end{picture}
}
\end{center}
\caption{Feynman diagrams for the nucleon current. The gray blob
represents the spin invariant for the coupled quark pair
in the effective Lagrangian, Eq.
(\ref{lag}). Diagram (1a) represents $j^+_{aN}$, Eq.(\ref{j+a}).
Diagram (1b) represents $j^+_{bN}$, Eq.(\ref{j+b}).
Diagram (1c) represents $j^+_{cN}$, Eq.(\ref{j+c}).
Diagram (1d) represents $j^+_{dN}$, Eq.(\ref{j+d}).
}
\label{fig1}
\end{figure}
\newpage
\vskip -1cm
\postscript{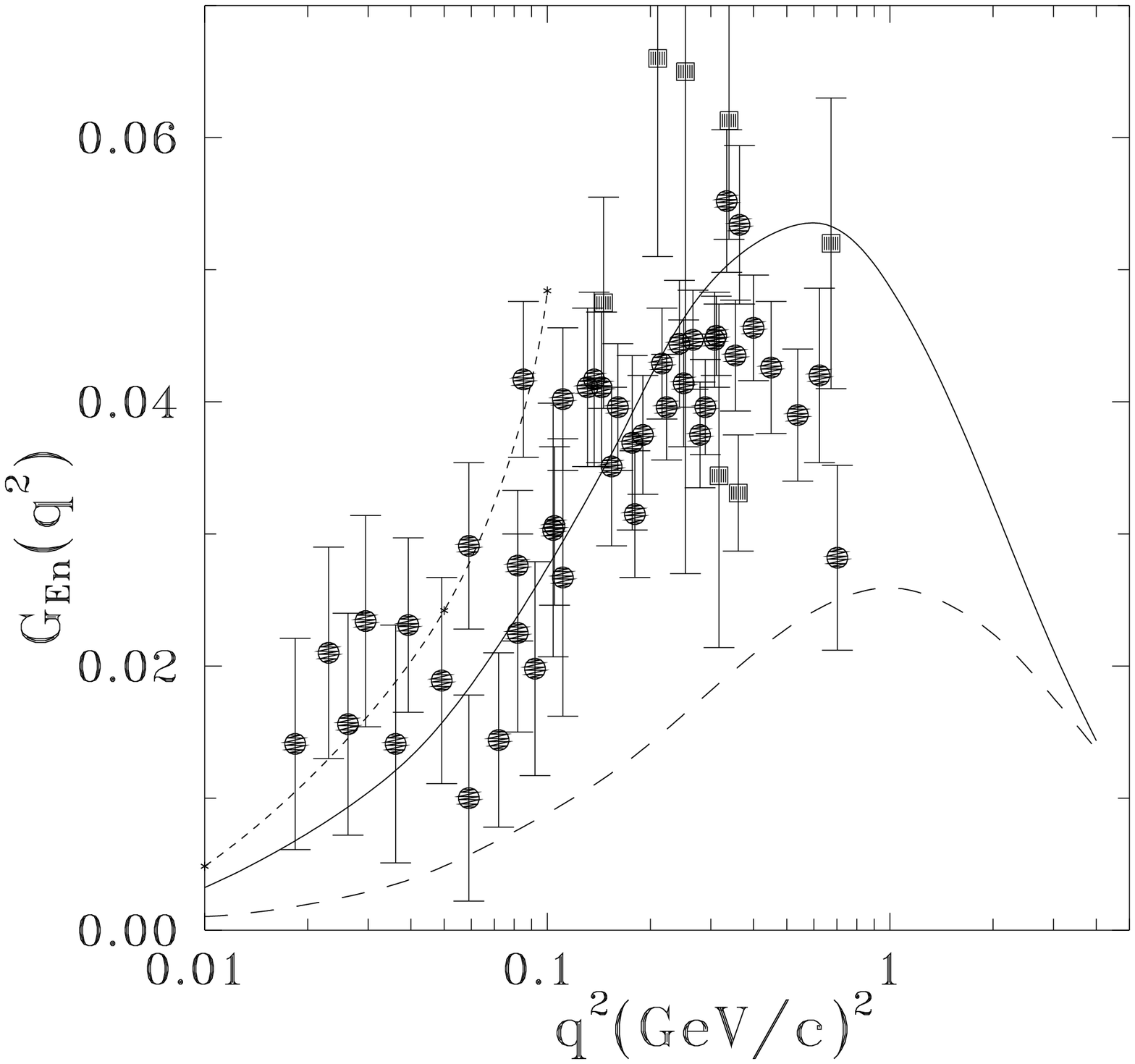}{.5}    
{\bf Fig. 2 } Neutron electric form factor as a function of the momentum
transfer $q^2=-Q^2$. Results for $\alpha=1$, solid line; for
$\alpha=1/2$, dashed line. The short-dashed line represents the curve
$G_{En}(Q^2)\approx \frac16<r^2>_n Q^2$ with the experimental value of
$<r^2>_n=-0.113 \ {\rm fm}^2$\cite{kop} .
The full circles are the experimental data from
Ref.\cite{plat} and the full squares from Ref.\cite{eden}.
\newpage
\vskip -1cm  
\postscript{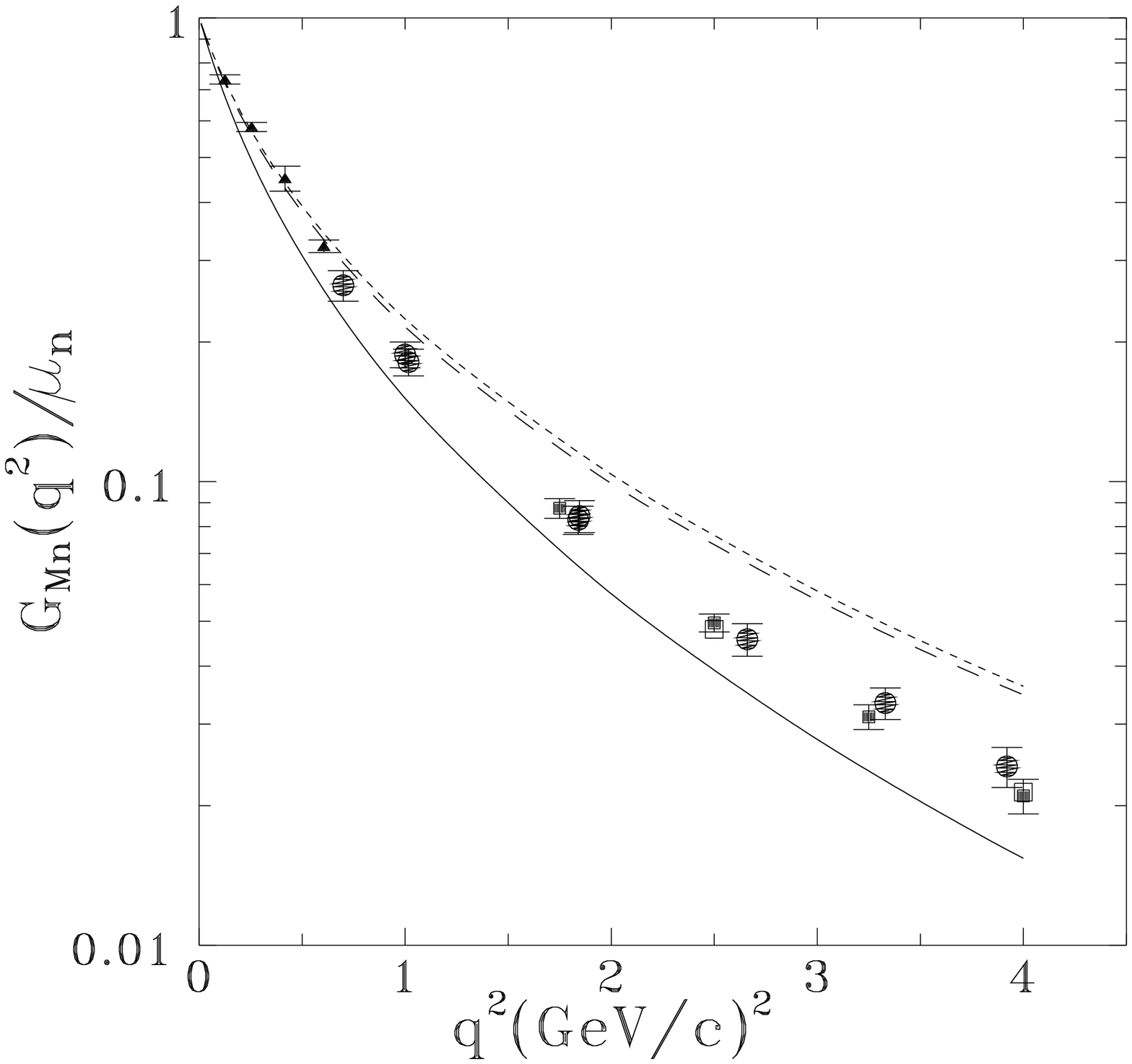}{0.5}    
{\bf Fig. 3 } Neutron magnetic form-factor  
as a function of the momentum transfer $q^2=-Q^2$. 
Model results  for $\alpha=1$, solid line; for
$\alpha=1/2$, dashed line; for $\alpha=0$, short-dashed line.
Experimental data from Ref.\cite{alb}, full circles;
from Ref.\cite{rock}, empty squares;
from Ref.\cite{lung}, full squares; 
from Ref.\cite{bruins}, full triangles. 

\newpage
\vskip -1cm 
\postscript{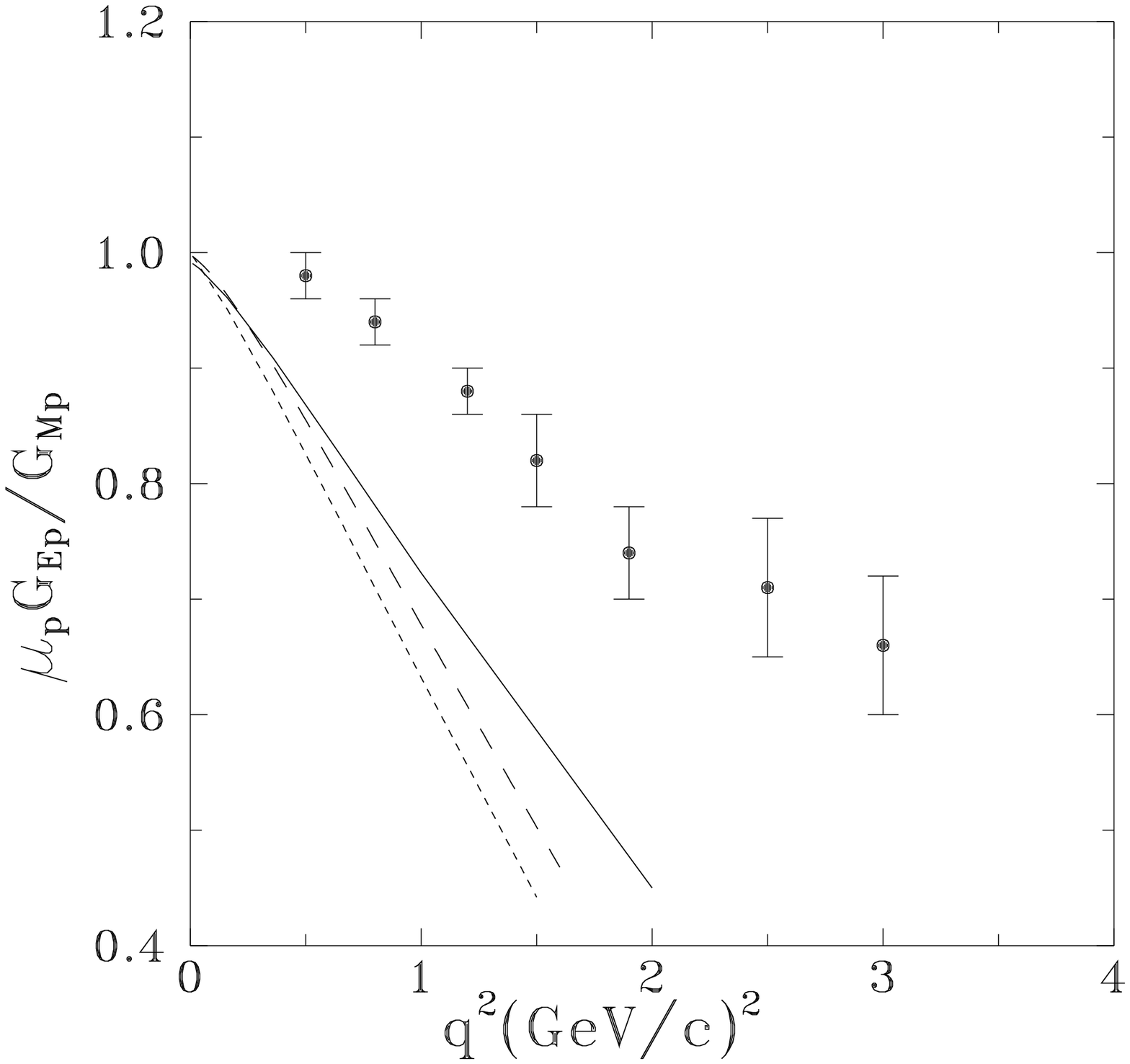}{0.5}    
{\bf Fig. 4 } 
Proton form-factor ratio $\mu_pG_{Ep}/G_{Mp}$ as a function
of momentum transfer. Theoretical curves labeled as in figure 3.
The experimental data comes from Ref.\cite{ran}.

\end{document}